\documentclass[aps,nofootinbib,twocolumn,prd,eqsecnum,showpacs,showkeys,preprintnumbers,altaffilletter]{revtex4-1}
\usepackage[caption=false]{subfig}
\usepackage{graphicx}
\usepackage{amsmath}
\usepackage{amsfonts}
\usepackage{amssymb}
\usepackage{color}
\usepackage{bm}
\usepackage{mathrsfs}
\usepackage{epstopdf}
\usepackage{url}
\usepackage{footnote}
\usepackage{textcomp}

\makeatletter
\newcommand*{\rom}[1]{\expandafter\@slowromancap\romannumeral #1@}
\makeatother

\begin{document}

\title{Modified Eddington-inspired-Born-Infeld Gravity with a Trace Term}
\author{Che-Yu Chen $^{1,2,3}$}
\email{b97202056@ntu.edu.tw}
\author{Mariam Bouhmadi-L\'{o}pez$^{4,5,6,7}$}
\email{{\mbox{mbl@ubi.pt. On leave of absence from UPV and IKERBASQUE.}}}
\author{Pisin Chen $^{1,2,3,8}$}
\email{pisinchen@phys.ntu.edu.tw}
\date{\today}

\affiliation{
${}^1$Department of Physics, National Taiwan University, Taipei, Taiwan 10617\\
${}^2$LeCosPA, National Taiwan University, Taipei, Taiwan 10617\\
${}^3$Graduate Institute of Astrophysics, National Taiwan University, Taipei, Taiwan 10617\\
${}^4$Departamento de F\'{i}sica, Universidade
da Beira Interior, 6200 Covilh\~a, Portugal\\
${}^5$Centro de Matem\'atica e Aplica\c{c}\~oes
da Universidade da Beira Interior (CMA-UBI), 6200 Covilh\~a, Portugal\\
${}^6$Department of Theoretical Physics, University of the Basque Country
UPV/EHU, P.O. Box 644, 48080 Bilbao, Spain\\
${}^7$IKERBASQUE, Basque Foundation for Science, 48011, Bilbao, Spain\\
${}^8$Kavli Institute for Particle Astrophysics and Cosmology, SLAC National Accelerator Laboratory, Stanford University, Stanford, CA 94305, U.S.A.
}

\begin{abstract}
In this paper, a modified Eddington-inspired-Born-Infeld (EiBI) theory with a pure trace term $g_{\mu\nu}R$ being added to the determinantal action is analysed from a cosmological point of view. It corresponds to the most general action constructed from a rank two tensor that contains up to first order terms in curvature. This term can equally be seen as a conformal factor multiplying the metric $g_{\mu\nu}$. This very interesting type of amendment has not been considered within the Palatini formalism despite the large amount of works on the Born-Infeld-inspired theory of gravity. This model can provide smooth bouncing solutions which were not allowed in the EiBI model for the same EiBI coupling. Most interestingly, for a radiation filled universe there are some regions of the parameter space that can naturally lead to a de Sitter inflationary stage without the need of any exotic matter field. Finally, in this model we discover a new type of cosmic ``quasi-sudden" singularity, where the cosmic time derivative of the Hubble rate becomes very large but finite at a finite cosmic time.
\end{abstract}

\keywords{modified theories of gravity, early time universe, Eddington-inspired-Born-Infeld theories, cosmic singularities}
\pacs{98.80.-k, 98.80.Jk, 04.50.Kd, 04.20.Dw}

\maketitle

\section{Introduction}\label{introduction}
Undeniably, Einstein's theory of general relativity (GR) has been an extremely successful theory for around a century \cite{gravitation}. However, the theory is expected to break down at some points at very high energies where quantum effects are expected to become crucial, such as in the past expansion of the Universe where GR predicts a big bang singularity \cite{largescale}. This is one of the motivations for looking for possible modified theories of gravity, which are hoped to not only be able to preserve the huge achievements of GR, but also to shed some light on smoothing the singularities predicted in GR. Such theories could be seen as effective/phenomenological approaches of a more fundamental quantum theory of gravity.

In 1934 Born and Infeld proposed a non-linear action for classical electrodynamics, which is characterised by its grand success in solving the divergence of the self-energy of point-like charges \cite{Born:1934gh}. This action for electromagnetism has an elegant determinantal structure which reads:
\begin{equation}
\mathcal{S}_{BI}=\frac{1}{\kappa}\int d^4x\sqrt{|g_{\mu\nu}+\kappa \mathcal{F}_{\mu\nu}|},
\label{BI action}
\end{equation}
with $g_{\mu\nu}$ and $\mathcal{F}_{\mu\nu}$ being the metric tensor and the electromagnetic field strength. Here $\kappa$ is a dimensional constant. Note that the Born-Infeld action \eqref{BI action} recovers Maxwell 
action for small amplitudes.

Since the proposal of the Born-Infeld action, modified theories of gravity with a Born-Infeld-inspired action, initiated by the pioneering work of Deser and Gibbons \cite{Deser:1998rj}, have received much attention; c.f. Refs.~\cite{Comelli:2004qr,Vollick:2003qp,Vollick:2005gc,Wohlfarth:2003ss,Nieto:2004qj,Comelli:2005tn}. These theories of gravity not only maintain the properties of GR for small curvatures, but provide various interesting deviations from GR at high curvature regimes. Most of these works start with a general gravitational action of the form
\begin{equation}
\mathcal{S}_{det}=\frac{1}{\kappa}\int d^4x\sqrt{|g_{\mu\nu}+\kappa \mathcal{G}_{\mu\nu}(R,R_{\alpha\beta},R_{\mu\nu\alpha\beta})|},
\label{det action}
\end{equation}
where $\mathcal{G}_{\mu\nu}$ represents a linear Ricci term $R_{\mu\nu}$, plus higher order curvature terms containing $R$, $R_{\mu\nu}$, and $R_{\mu\nu\alpha\beta}$. Rather than amending the higher curvature terms in $\mathcal{G}_{\mu\nu}$, we can also modify the action by multiplying the metric term in the determinant by a factor $1+\alpha_1 R$, where $\alpha_1$ is a constant with length square dimensions. This approach does not lead to a violation of the requirements in the low curvature limit, i.e. we can still recover GR at low energies. Actually, a gravitational action with this pure trace term has been considered in Ref.~\cite{Comelli:2004qr} within a pure metric formalism. The theory thus inevitably suffers from the presence of troublesome fourth order derivatives in the field equations or ghost instabilities \cite{Deser:1998rj}. 

In order to keep the theory free from aforementioned problems, alternative theories formulated within the Palatini formalism and teleparallel representation have been widely studied in Refs.~\cite{Vollick:2003qp,Vollick:2005gc,Banados:2010ix,Ferraro:2006jd,Ferraro:2008ey,Fiorini:2009ux,Ferraro:2009zk,Fiorini:2013kba}. For example, a theory constructed upon the Palatini approach, which is dubbed Eddington-inspired-Born-Infeld theory (EiBI) (see Ref.~\cite{Banados:2010ix}), has recently attracted a lot of attention and has been studied from both astrophysical and cosmological points of view \cite{Potapov:2014iva,Izmailov:2015xsa,Tamang:2015tmd,Bouhmadi-Lopez:2013lha,Scargill:2012kg,Avelino:2012ue,EscamillaRivera:2012vz,Yang:2013hsa,Du:2014jka,Wei:2014dka,Delsate:2012ky,Pani:2011mg,Pani:2012qb,Casanellas:2011kf,Avelino:2012ge,Avelino:2012qe,Harko:2013wka,Harko:2013aya,Sham:2013cya,Makarenko:2014lxa,Makarenko:2014nca,Odintsov:2014yaa,Pani:2012qd,Bouhmadi-Lopez:2014jfa,Makarenko:2014cca,Shaikh:2015oha,Jana:2015cha,Sotani:2015tya,Cho:2014xaa,Sotani:2014lua,Bouhmadi-Lopez:2014tna,Cho:2012vg,Cho:2013pea,Cho:2013usa}. The EiBI theory is shown to be able to cure the big bang singularity for a radiation dominated universe through a loitering effect\footnote{The Universe starts its evolution with a minimum size and at an infinite cosmic time in the past, before it enters the standard GR expansion \cite{Banados:2010ix}.} or a bounce\footnote{A bouncing universe is a universe with a minimum or a maximum scale factor such that after a contracting phase, an expanding phase happens or the other way around. In this kind of model, the big bang is substituted by a bounce.} in the past, with the coupling constant $\kappa$ being positive or negative, respectively. The ability of the theory to smooth other cosmological singularities in a phantom dominated universe was also studied in Refs.~\cite{Bouhmadi-Lopez:2013lha,Bouhmadi-Lopez:2014jfa}. Interestingly, in Refs.~\cite{Makarenko:2014lxa,Odintsov:2014yaa,Makarenko:2014cca} the authors showed that the bouncing solutions for negative $\kappa$ are robust against the changes of the Lagrangian through an additional $f(R)$ term or some functional extensions (see as well Ref.~\cite{Jimenez:2014fla} for another generalized gravitational theory related to massive gravity and Ref.~\cite{Jimenez:2015caa} for the tensorial perturbations of a further generalized gravitational theory within the Palatini formalism.).
However, it should be stressed that the amendments through the addition of a pure trace term to the determinantal action have never been considered so far. Besides, the EiBI theory with a negative coupling constant $\kappa$ was also shown to suffer from instability problems due to the imaginary effective sound speed \cite{Avelino:2012ge}.

On the other hand, a recently proposed determinantal gravity formulated within the teleparallel representation was shown to be able to cure the big bang singularity in the past evolution of the Universe through a de Sitter inflationary phase \cite{Fiorini:2013kba}. Considering the widest generalization, the author added a pure trace term into the Lagrangian of the form of $g_{\mu\nu}\boldsymbol{T}$, where $\boldsymbol{T}$ is the Weitzenb\"{o}ck invariant \cite{tele}. In our previous work, we exhibited that this theory contains cosmological singularities for some parameters of the model, including the emergence of some cosmological singularities from purely geometrical effects (without the need of exotic matter) \cite{Bouhmadi-Lopez:2014tna}. 

As far as we know, the gravitational actions with a pure trace term added to the Born-Infeld determinantal structure have never been considered within the Palatini approach in the literature. Furthermore, we expect the emergence of interesting cosmological solutions with the addition of a pure trace term because it is expected in the teleparallel version. Based on these motivations, in this work we will consider a modified EiBI theory with a pure trace term added to the determinantal action, and analyse its cosmological implications. For simplicity, in this work we will assume a homogeneous and isotropic universe filled with a perfect fluid with a constant equation of state. Because the field equations are complicated, we will follow a method similar to that used in Ref.~\cite{Makarenko:2014lxa} to demonstrate the results graphically. 

In this paper, we will follow Ref. \cite{Nojiri:2005sx} to characterize the cosmological singularities by the behavior of the Hubble rate and its cosmic time derivative at the singular event: \footnote{For an alternative classification of cosmological singularities see Refs.~\cite{FernandezJambrina:2004yy,FernandezJambrina:2006hj,Fernandez-Jambrina:2014sga}.}

\begin{itemize}

\item A big rip singularity takes place at a finite cosmic time with an infinite scale factor, where the Hubble parameter and its cosmic time derivative diverge \cite{Starobinsky:1999yw,Caldwell:2003vq,Caldwell:1999ew,Carroll:2003st,Chimento:2003qy,Dabrowski:2003jm,GonzalezDiaz:2003rf,GonzalezDiaz:2004vq}.

\item A sudden singularity takes place at a finite cosmic time with a finite scale factor, where the Hubble parameter remains finite but its cosmic time derivative diverges \cite{Barrow:2004xh,Gorini:2003wa,Nojiri:2005sx}.

\item A big freeze singularity takes place at a finite cosmic time with a finite scale factor, where the Hubble parameter and its cosmic time derivative diverge \cite{BouhmadiLopez:2007qb,BouhmadiLopez:2006fu,Nojiri:2005sx,Nojiri:2004pf,Nojiri:2005sr}.

\item A type \rom{4} singularity takes place at a finite cosmic time with a finite scale factor, where the Hubble parameter and its cosmic time derivative remain finite, but higher cosmic time derivatives of the Hubble parameter still diverge \cite{BouhmadiLopez:2006fu,Nojiri:2005sx,Nojiri:2004pf,Nojiri:2005sr,Nojiri:2008fk,Bamba:2008ut}.

\item A little rip event takes place at an infinite cosmic time with an infinite scale factor, where the Hubble rate and its cosmic time derivative diverge \cite{Ruzmaikina1970,Nojiri:2005sx,Stefancic:2004kb,BouhmadiLopez:2005gk,Bouhmadi-Lopez:2013nma,Frampton:2011sp,Brevik:2011mm}.

\item A little sibling of the big rip takes place at an infinite cosmic time with an infinite scale factor, where the Hubble rate diverges, but its cosmic time derivative remains finite \cite{Bouhmadi-Lopez:2014cca}.

\end{itemize}

Our results are clearly shown in Table~\ref{summary} where we compare them with the original EiBI model \cite{Banados:2010ix}. As can be seen the model we are proposing can provide smooth bouncing solutions which were not allowed in the EiBI model for the same EiBI coupling ($\kappa>0$). Most interestingly, for a radiation filled universe there are some regions of the parameter space that can naturally lead to a de Sitter inflationary stage without the need of any exotic matter field. Finally, in this model we discover a new type of cosmic ``quasi-sudden" singularity, where the cosmic time derivative of the Hubble rate becomes very large but finite at a finite cosmic time.

This paper is outlined as follows. In section~II, we briefly introduce the basis of the modified EiBI theory with the addition of a pure trace term, including its action, field equations, and the low curvature limits of the theory. In section~III, we assume a homogeneous and isotropic universe filled with a perfect fluid with a constant equation of state, then follow a similar approach to that used in Ref.~\cite{Makarenko:2014lxa} to derive a parametric Friedmann equation. In section~IV, we exhibit the evolution of the Universe by graphically showing the Hubble rate as a function of the energy density under different assumptions of the parameters characterising the theory. To analyse the evolution of the Universe at the very early time, we then confine ourselves to a radiation dominated universe in our analysis of the modified EiBI theory. We finally present our conclusions in section~V.

\section{Proposed model: action and field equations}\label{sectII}

As mentioned in the introduction, in this paper we will add a pure trace term, which takes the form of $g_{\mu\nu}R$, to the EiBI determinantal Lagrangian. This term can equally be seen as a conformal factor multiplying the metric $g_{\mu\nu}$. Therefore, the action of this generalized EiBI theory is
\begin{equation}
\mathcal{S}=\frac{1}{\kappa}\int d^4x\Big[\sqrt{|g_{\mu\nu}+\kappa F_{\mu\nu}|}-\lambda\sqrt{-g}\Big]+S_m,
\label{action}
\end{equation}
where
\begin{equation}
F_{\mu\nu}=\alpha R_{\mu\nu}(\Gamma)+\beta g_{\mu\nu}R.
\label{F}
\end{equation}
The theory is formulated within the Palatini formalism, in which the metric $g_{\mu\nu}$ and the connection $\Gamma$ are treated as independent variables. In addition, $R_{\mu\nu}(\Gamma)$ is chosen to be the symmetric part of the Ricci tensor and the connection is also assumed to be torsionless.
Note that $g$ is the determinant of the metric and $S_m$ stands for the matter Lagrangian, where matter is assumed to be coupled covariantly to the metric $g$ only. $\alpha$, $\beta$ and $\lambda$ are dimensionless constants. The parameter $\kappa$ is a constant with inverse dimensions to that of a cosmological constant (in this paper, we will work with Planck units $8\pi G=1$ and set the speed of light to $c=1$). 

In the low energy limit ($\kappa\rightarrow 0$), the gravitational action \eqref{action} becomes
\begin{eqnarray}
\mathcal{S}_g&\approx &\frac{1}{2}\int d^4 x\sqrt{-g}\Big[(\alpha+4\beta)R-2\Lambda\nonumber\\
&+&(\frac{1}{4}\alpha^2+\alpha\beta+2\beta^2)\kappa R^2\nonumber\\
&-&\frac{1}{2}\alpha^2 \kappa R_{\mu\nu}R^{\mu\nu}+O^3(R)\Big],
\label{low energy action}
\end{eqnarray}
where the effective cosmological constant is defined by $\Lambda\equiv(\lambda-1)/\kappa$. Therefore, the dimensionless constants $\alpha$ and $\beta$ should satisfy:
\begin{equation}
\alpha+4\beta=1
\label{criteria}
\end{equation}
to ensure the recovering of Einstein GR at the low curvature limit. Moreover, it can be easily seen that the EiBI theory is regained when $\alpha=1$ and $\beta=0$. On the other hand, this theory  becomes an $R^2$ theory with its gravitational action part being
\begin{equation}
S_g|_{\alpha=0}=\frac{1}{2}\int d^4x\sqrt{-g}\Big(R+\frac{\kappa}{8}R^2-2\Lambda\Big),
\end{equation}
when $\alpha=0$ and $\beta=1/4$ \cite{Meng:2003bk}. Note that this constitutes the sole of $f(R)$-like action that can be derived from the determinantal structure, which is also valid for $f(\boldsymbol{T})$-like action in the teleparallel representation \cite{Fiorini:2013kba}. Therefore, the dimensionless constants $\alpha$ and $\beta$ in this theory can be used to quantify the extent of the interpolation between the Palatini $R^2$ theory and the EiBI theory.

Actually, one can also add the so called zeroth order curvature term; i.e., $\gamma g_{\mu\nu}$, to the determinant based on the structural completeness. However, This additional term can be rescaled by a conformal transformation $g_{\mu\nu}\rightarrow(1+\gamma)g_{\mu\nu}$ and then can be absorbed into the cosmological constant term. In this sense this additional term is not expected to affect our results significantly, especially at the high energy regime in which the influence caused by high curvature terms is dominant. In fact, one can easily see from the gravitational action that the higher order curvatures term will dominate over the zeroth order term when curvature gets large. Therefore, we will omit this possible additional term in this work.

Within the Palatini approach we are assuming here, we have to vary the action \eqref{action} with respect to the metric and the connection independently to derive the complete field equations. After varying the action with respect to $g_{\mu\nu}$, we derive the first field equation
\begin{eqnarray}
&&\frac{\sqrt{-q}}{\sqrt{-g}}q^{\mu\nu}(1+\kappa\beta R)-\lambda g^{\mu\nu}\nonumber\\
&-&\frac{\sqrt{-q}}{\sqrt{-g}}\kappa\beta q^{\alpha\beta}g_{\alpha\beta}g^{\mu\rho}g^{\nu\sigma}R_{\rho\sigma}=-\kappa T^{\mu\nu},
\label{eq1}
\end{eqnarray}
where $q_{\mu\nu}\equiv g_{\mu\nu}+\kappa F_{\mu\nu}$ and $q^{\mu\nu}$ is the inverse of $q_{\mu\nu}$. $T^{\mu\nu}$ stands for the energy momentum tensor. Because the matter is assumed to be coupled covariantly to the metric $g$ only, the energy momentum tensor is conserved like in GR \cite{Banados:2010ix}.

The second field equation can be obtained by varying the action \eqref{action} with respect to the connection
\begin{equation}
\nabla_\nu\Big[\sqrt{-q}(\alpha q^{\mu\nu}+\beta q^{\alpha\beta}g_{\alpha\beta}g^{\mu\nu})\Big]=0.
\label{eq2}
\end{equation}
Note that the covariant derivative $\nabla_\nu$ is defined through the connection $\Gamma$.

\section{Modified EiBI gravity: A parametric Friedmann equation}

To analyse the behavior of the cosmological solutions in the generalized theory defined in Eqs.~\eqref{action} and \eqref{F}, we follow an approach similar to the one proposed in Ref.~\cite{Makarenko:2014lxa} to our model. More precisely, we will rewrite the field equations in an algebraic form and express the quantities of interest using a single variable $x$, then we will represent the behavior of the cosmological solutions graphically. We assume $\hat{q}$ and $\hat{q}^{-1}$ denoting $q_{\mu\nu}$ and its inverse $q^{\mu\nu}$, respectively. From now on, a hat will denote a tensor without making explicit reference to the tensor components. We further define $\hat{\Omega}=\hat{g}^{-1}\hat{q}$ and $\hat{\Omega}^{-1}=\hat{q}^{-1}\hat{g}$ to rewrite the first field equation \eqref{eq1} as follows
\begin{eqnarray}
&&|\hat{\Omega}|^{\frac{1}{2}}\hat{\Omega}^{-1}(1+\kappa\beta R)-|\hat{\Omega}|^{\frac{1}{2}}Tr(\hat{\Omega}^{-1})\kappa\beta(\hat{g^{-1}R})-\lambda\hat{I}\nonumber\\
&=&-\kappa\hat{T},
\label{field eq ome}
\end{eqnarray}
where $(\hat{g^{-1}R})\equiv g^{\mu\alpha}R_{\alpha\nu}(\Gamma)$ and $\hat{T}\equiv T^{\mu\alpha}g_{\alpha\nu}$. Note that $\hat{I}$ is the identity matrix and $Tr{\hat{A}}$ denotes the trace of a matrix $\hat{A}$. 

After taking a trace of both sides of Eq.~\eqref{field eq ome}, we get
\begin{equation}
|\hat{\Omega}|^{\frac{1}{2}}Tr(\hat{\Omega}^{-1})=4\lambda-\kappa T.
\label{Tbar}
\end{equation}
Moreover, according to the definition of $\hat{q}$ and $\hat{\Omega}$ and Eq.~\eqref{criteria} we have
\begin{eqnarray}
Tr{\hat{\Omega}}&=&4(1+\kappa\beta R)+\kappa\alpha R=4+\kappa R,\label{TrO}\\
(\hat{g^{-1}R})&=&\frac{\hat{\Omega}-(1+\kappa\beta R)\hat{I}}{\alpha\kappa}.\label{gR}
\end{eqnarray}
Note that equation \eqref{gR} is valid only for $\alpha\neq 0$. Combining Eqs.~\eqref{field eq ome}, \eqref{Tbar}, \eqref{TrO} and \eqref{gR}, and after some algebra, we obtain
\begin{equation}
\frac{\alpha(\alpha+\beta Tr\hat{\Omega})}{Tr(\hat{\Omega}^{-1})}\hat{\Omega}^{-1}-\beta\hat{\Omega}+\frac{\alpha\hat{T}}{\bar{T}}+\Big[\beta(\alpha+\beta Tr\hat{\Omega})-\frac{\alpha\lambda}{\kappa\bar{T}}\Big]\hat{I}=0,
\label{eq3}
\end{equation}
where $\bar{T}\equiv4\lambda/\kappa-T$.

To analyse the solutions within a cosmological scale; i.e., we assume the cosmological principle, we first assume that the Universe is homogeneous and isotropic at large scale and that it is filled with an effective perfect fluid with energy density $\rho$ and pressure $p$. Then $\hat{\Omega}$ becomes a diagonal tensor with 
\begin{eqnarray}
\Omega^0_0&\equiv&\omega_1,\nonumber\\
\Omega^i_j&\equiv&\omega_2\delta^i_j.
\label{omecom}
\end{eqnarray}
Therefore, the non-vanishing components of Eq.~\eqref{eq3} read,
\begin{eqnarray}
&&\frac{\alpha[\alpha+\beta(\omega_1+3\omega_2)]\omega_2}{\omega_2+3\omega_1}-\beta\omega_1-\frac{\alpha\rho}{\bar{T}}\nonumber\\
&=&\frac{\alpha\lambda}{\kappa\bar{T}}-\beta[\alpha+\beta(\omega_1+3\omega_2)],
\label{rho1}
\end{eqnarray}
and
\begin{eqnarray}
&&\frac{\alpha[\alpha+\beta(\omega_1+3\omega_2)]\omega_1}{\omega_2+3\omega_1}-\beta\omega_2+\frac{\alpha p}{\bar{T}}\nonumber\\
&=&\frac{\alpha\lambda}{\kappa\bar{T}}-\beta[\alpha+\beta(\omega_1+3\omega_2)].
\label{p1}
\end{eqnarray}

Next, we introduce a constant equation of state $w$ for the perfect fluid, i.e. it satisfies $p=w\rho$. Combining Eqs.~\eqref{rho1} and \eqref{p1}, the energy density as functions of $\omega_1$ and $\omega_2$ can be written as:
\begin{equation}
\kappa\rho =\frac{4\lambda X(\omega_2-\omega_1)}{\alpha(1+w)-X(\omega_2-\omega_1)(1-3w)},
\label{rho}
\end{equation}
where
\begin{equation}
X=\frac{\alpha[\alpha+\beta(\omega_1+3\omega_2)]}{\omega_2+3\omega_1}+\beta.
\end{equation}

Then, we introduce a dimensionless parameter $x$, which satisfies 
\begin{eqnarray}
\omega_1=x^3|\hat{\Omega}|^{\frac{1}{4}}\nonumber\\
\omega_2=\frac{|\hat{\Omega}|^{\frac{1}{4}}}{x}.
\end{eqnarray}
Therefore, according to Eq.~\eqref{TrO}, we have 
\begin{equation}
x^3+\frac{3}{x}=4z,
\label{zde}
\end{equation}
where $4z=(4+\kappa R)/|\hat{\Omega}|^{\frac{1}{4}}$.

Next, by adding Eq.~\eqref{p1} to Eq.~\eqref{rho1} after being multiplied by $w$, the terms involving $\rho$ are cancelled out, and an algebraic equation for $x$ and $|\hat{\Omega}|^{\frac{1}{4}}$ is obtained:
\begin{eqnarray}
&&\frac{\alpha(\alpha+4\beta z|\hat{\Omega}|^{\frac{1}{4}})}{1+3x^4}(w+x^4)-\beta(wx^3+\frac{1}{x})|\hat{\Omega}|^{\frac{1}{4}}\nonumber\\
&&+\beta(\alpha+4\beta z|\hat{\Omega}|^{\frac{1}{4}})(w+1)=\frac{\alpha\lambda(w+1)x^3}{(1+3x^4)|\hat{\Omega}|^{\frac{1}{4}}}.\nonumber\\
\end{eqnarray}
After some rearrangements we derive a quadratic equation for $|\hat{\Omega}|^{\frac{1}{4}}$ which reads
\begin{equation}
R_1|\hat{\Omega}|^{\frac{1}{2}}+R_2|\hat{\Omega}|^{\frac{1}{4}}+R_3=0,
\end{equation}
where
\begin{eqnarray}
R_1&=&4\alpha\beta z(w+x^4)-\beta(w x^3+\frac{1}{x})(1+3x^4)\nonumber\\
&+&4{\beta}^2 z(w+1)(1+3x^4),\\
R_2&=&{\alpha}^2(w+x^4)+\alpha\beta(w+1)(1+3x^4),\\
R_3&=&-\alpha\lambda(w+1)x^3.
\end{eqnarray}
The solution to this quadratic equation, expressed in terms of $x$, reads
\begin{eqnarray}
|\hat{\Omega}|^{\frac{1}{4}}=
\begin{cases}
\frac{-R_2\pm\sqrt{{R_2}^2-4R_1R_3}}{2R_1}, & R_1\neq 0 \\
-\frac{R_3}{R_2}, & R_1=0.
\end{cases}
\label{om4}
\end{eqnarray}
In addition, after factoring $R_1(x)$ we find that if $\beta=0$ (EiBI limit) or $w=1/3$ (radiation domination), $R_1(x)$ vanishes, which means the second equation in Eq.~\eqref{om4} is valid. Furthermore, it should be stressed that the approach mentioned above can not be applied to the case in which $\alpha=0$ (see Eq.~\eqref{gR}). Actually, $x$ is fixed to be $x=1$ in this case, thus $x$ is no longer a changing variable. However, this fact is not a real problem because one can easily derive the cosmological solutions for a pure $R^2$ action without the need of the approach we are following \cite{Meng:2003bk}.

We have now derived the expression of the energy density as a function of $x$ in Eq.~\eqref{rho}. If we can further express the Hubble rate as a function of $x$, the graphical relationship between the Hubble rate and the energy density can be completed. 

As was already assumed at the beginning of this section, we focus on a cosmological symmetry and choose a spatially flat Friedmann-Lema\^itre-Robertson-Walker (FLRW) space-time:
\begin{equation}
ds^2=-dt^2+a^2(t)(dx^2+dy^2+dz^2),
\label{metric}
\end{equation}
where $t$ is the cosmic time and $a(t)$ is the scale factor.

We then define two tensors $\hat{\Sigma}$ and $h_{\mu\nu}$ such that $\sqrt{-q}\hat{\Sigma}q^{\mu\nu}=\sqrt{-h}h^{\mu\nu}$, and $\hat{\Sigma}=\alpha\hat{I}+\beta Tr({\hat{\Omega}}^{-1})\hat{\Omega}$. Note that $h^{\mu\nu}$ denotes the inverse of $h_{\mu\nu}$. According to the field equation \eqref{eq2}, the tensor $h_{\mu\nu}$ is the auxiliary metric which is compatible with the physical connection of this theory. After some calculations, we obtain $|q||\Sigma|=|h|$. Inserting it back we have $h^{\mu\nu}=\hat{\Sigma}\hat{q}^{-1}/{|\hat{\Sigma}|}^{1/2}$, and $h_{\mu\nu}=|\hat{\Sigma}|^{1/2}\hat{g}\hat{\Omega}\hat{\Sigma}^{-1}$. The non-vanishing components of $\hat{\Sigma}$ and $h_{\mu\nu}$ read
\begin{eqnarray}
\Sigma^0_0\equiv\sigma_1&=&\alpha+\beta(1+3x^4),\nonumber\\
\Sigma^i_j\equiv\sigma_2\delta^i_j&=&\Big[\alpha+\beta(\frac{1}{x^4}+3)\Big]\delta^i_j,
\label{sigmacom}
\end{eqnarray}
and
\begin{eqnarray}
h_{00}&=&-\sqrt{\frac{{\sigma_2}^3}{\sigma_1}}\omega_1\equiv -h_1,\nonumber\\
h_{ij}&=&\sqrt{\sigma_1\sigma_2}\omega_2\delta_{ij}a^2\equiv h_2\delta_{ij}a^2.
\label{hcom}
\end{eqnarray}

Once we have obtained the components of the auxiliary metric $h_{\mu\nu}$ which is compatible with the physical connection of the theory, we can derive the components of the connection and the Ricci tensor. After some lengthy calculations, we obtain
\begin{eqnarray}
R_{00}+\frac{h_1}{h_2a^2}R_{ij}\delta^{ij}&=&\frac{3}{2}\Big(\frac{\dot{h_2}}{h_2}+2H\Big)^2\nonumber\\
&=&\frac{3}{2}\Big[2-\frac{3}{h_2}\frac{d h_2}{d\rho}\rho(1+w)\Big]^2H^2,\nonumber\\
\end{eqnarray}
where $H=\dot{a}/a$ is the Hubble rate and the dot denotes the cosmic time derivative. Note that the conservation equation $\dot{\rho}=-3H(\rho+p)$ has been applied in the above equation. Finally, we arrive at the expression of $H^2$ in terms of the variable $x$:
\begin{eqnarray}
&&\frac{1}{2}\kappa H^2\nonumber\\
&=&\frac{\alpha+|\hat{\Omega}|^{\frac{1}{4}}(4\beta z-x^3)+3\frac{\sigma_2}{\sigma_1}[|\hat{\Omega}|^{\frac{1}{4}}(x^3-4\beta zx^4)-\alpha x^4]}{3\alpha[2-\frac{3}{h_2}\frac{d h_2}{d\rho}\rho(1+w)]^2},\nonumber\\
\label{mfe}
\end{eqnarray}
In the next section, we will combine Eqs.~\eqref{rho} and \eqref{mfe} together with the definitions \eqref{omecom}, \eqref{zde}, \eqref{sigmacom} and \eqref{hcom} to analyse the past/future asymptotic behavior of a FLRW universe in this type of model when filled with a perfect fluid with constant $w$. Eqs.~\eqref{rho} and \eqref{mfe} correspond to a parametric Friedmann equation being $x$ the free parameter.

\section{The Hubble rate in an expanding universe}

In the previous section, we derive the expressions of $\rho$ (Eq.~\eqref{rho}) and $H^2$ (Eq.~\eqref{mfe}) as functions of a single variable $x$. Therefore, we can graphically obtain the representations of $\kappa H^2$ as a function of $\kappa\rho$ to exhibit the behaviors of the cosmological solutions of interest. From now on, we will assume a vanishing cosmological constant to simplify the analysis, that is, we assume $\lambda=1$.

\subsection{The original EiBI theory: $\beta=0$}

\begin{figure}[!h]
\graphicspath{{fig/}}
\includegraphics[scale=0.8]{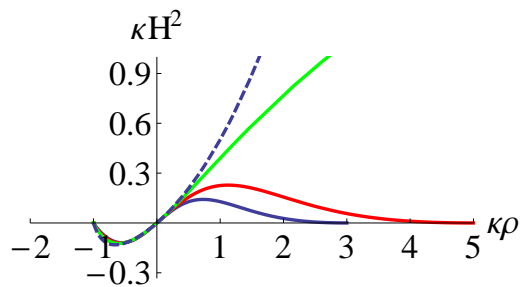}
\caption{Graphical representation of $\kappa H^2$ as a function of $\kappa\rho$ in the EiBI theory in which $\beta=0$. Different curves represent different equation of state $w$. The solid blue, red, green, and dashed blue curves show the plots for $w=1/3$, $1/5$, $0$, and $w=-1/3$, respectively.}
\label{EiBI}
\end{figure}

As a first glance, we consider the original EiBI theory in which $\beta=0$ and $\alpha=1$. The representations of $\kappa H^2$ as a function of $\kappa\rho$ are shown in FIG.~\ref{EiBI}. One can see that the evolution of the energy density terminates at a bounce where $H^2=0$ and $dH/d\rho\neq 0$ at $|\kappa\rho|=1$ for $\kappa<0$. This bouncing solution is robust against the change of the equation of state $w$. However, if $\kappa>0$ it can be seen that the behavior of the Hubble parameter is highly sensitive to the choice of $w$. There are loitering solutions where $H^2\rightarrow 0$ and $dH/d\rho\rightarrow0$ at $\kappa\rho=1/w$ for $w>0$, and divergent solutions for $w\le 0$. Furthermore, it can also be easily seen that the behaviors of the different curves focus around $H^2=\rho/3$ when $\rho\approx 0$. This property is not a surprise because of the prior criteria shown in Eq.~\eqref{criteria}, and it can be affirmed in all results shown in the rest of this paper; i.e. we recover GR at low energies. Note that the results summarized in this subsection are compatible with those concluded in the literatures \cite{Banados:2010ix,Scargill:2012kg}. 

\subsection{Radiation dominated universe}
In this subsection, we analyse if the original loitering behaviors and the bouncing solutions within the EiBI theory can be altered with the addition of a pure trace term $g_{\mu\nu}R$ to the determinantal Lagrangian, i.e., $\beta\neq 0$, for a radiation dominated universe. The analysis could be easily extended to other equation of state but for simplicity we stick to a radiation dominated universe.

\subsubsection{$\beta\gtrsim0$}

\begin{figure}[!h]
\graphicspath{{fig/}}
\includegraphics[scale=0.8]{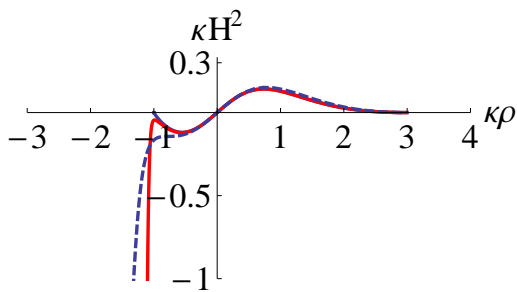}
\caption{Graphical representation of $\kappa H^2$ as a function of $\kappa\rho$ for a radiation dominated universe ($w=1/3$). Different curves represent different values of $\beta$. The solid blue, red, and dashed blue curves show the plots for $\beta=0$, $10^{-3}$, and $10^{-2}$, respectively.}
\label{rad1}
\end{figure}

We first consider the region in which $\beta$ is slightly larger than zero. One should be reminded that in Refs.~\cite{Makarenko:2014lxa,Odintsov:2014yaa,Makarenko:2014cca} the authors concluded that the bouncing solutions in the EiBI theory for negative $\kappa$ are robust against the amendment to the EiBI action through an additional $f(R)$ term or some functional extents. However, the situations are different in our model. One can see from FIG.~\ref{rad1} that the bouncing solutions for negative $\kappa$ are quite sensitive to the increase of $\beta$ from zero by even a small amount. More precisely, the asymptotic behavior of $H^2$ at large $\rho$ is $H^2\propto\rho$. This implies the occurrence of a big bang singularity in the past.

On the other hand, for positive $\kappa$ the loitering effect in the EiBI theory becomes a bounce in this model, in which $H^2\propto\delta\rho\equiv\rho-\rho_{\textrm{max}}$ with $\rho_{\textrm{max}}$ being the maximum energy density at the bounce. This gives a much regular behavior as compared with the asymptotic past behavior for $\kappa<0$.

\subsubsection{$0<\beta
\leq 1/4$}

\begin{figure}[!h]
\graphicspath{{fig/}}
\includegraphics[scale=0.8]{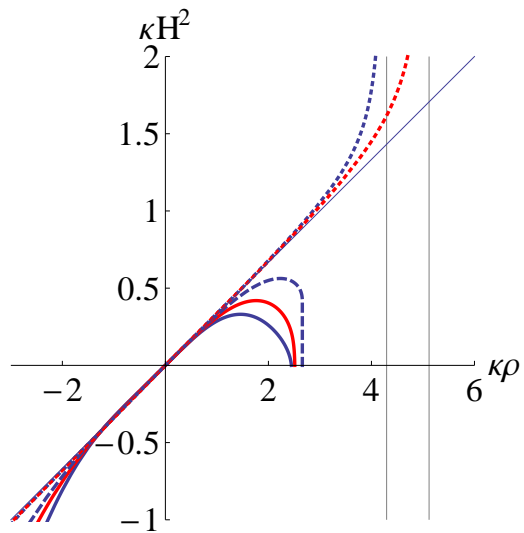}
\caption{Graphical representation of $\kappa H^2$ as a function of $\kappa\rho$ for a radiation dominated universe ($w=1/3$). Different curves represent different values of $\beta$. The solid blue, red, dashed blue, dotted blue, and dotted red curves show the representations for $\beta=1/10$, $3/25$, $7/50$, $1/5$, and $21/100$, respectively. The straight line crossing the origin represents $H^2=\rho/3$, which is the solution within $R+R^2$ gravity ($\beta=1/4$) for a radiation dominated universe. Note that the vertical lines (from left to right) indicate the maximum values of the energy density for $\beta=1/5$ and $21/100$, where the Hubble rate blows up, respectively. These singular events correspond to a big freeze singularity in the past.}
\label{rad2}
\end{figure}

As the value of $\beta$ increases and gets closer to $\beta\approx 1/4$, one can see from FIG.~\ref{rad2} that the behaviors of the big bang solutions gradually converge to those within the $R+R^2$ gravity ($\beta=1/4$), that is, $H^2=\rho/3$, for negative $\kappa$. On the other hand, if $\kappa>0$ the loitering behaviors within the EiBI theory can be substituted by other interesting cosmological solutions. For example, we find that for $\beta=1/10$ and $\beta=3/25$, the asymptotic behaviors of $H^2$, when $\rho$ approaches its maximum value $\rho_{\textrm{max}}$, become
\begin{equation}
H^2\propto\delta\rho,
\end{equation}
where $\delta\rho=\rho-\rho_{\textrm{max}}$. Combining it with the conservation equation $\rho\propto a^{-4}$, one can see that this event corresponds to a bounce in the past.

Furthermore, we have also found that the absolute value of $dH^2/d\rho$, which is proportional to $\dot{H}$ in this model as the energy momentum tensor is conserved, is a growing function of $\beta$. As $\beta$ approaches $\beta\approx\beta_{\star}=7/50$, $|dH^2/d\rho|$ gets very large at a finite past cosmic time. Therefore, this singular event can be regarded as a quasi-sudden singularity in the past on the sense that while $H$ is finite, $\dot{H}$ almost blows up in a finite past cosmic time. \footnote{We gave in the introduction the definition of the sudden singularity and the other cosmological singularities related to dark energy. In our case this singular event happens in the finite past of the Universe.}

However, if $\beta>\beta_{\star}$, i.e. larger than the value corresponding to a quasi-sudden singularity, the situation changes drastically. For $\beta=1/5$ and $21/100$, we find that the asymptotic behavior of $H^2$ reads
\begin{equation}
H^2\propto\delta\rho^{-2},
\end{equation}
when $\rho$ approaches $\rho_{\textrm{max}}$. Therefore, this event takes place at a finite scale factor and a finite cosmic time, with both $H$ and its cosmic time derivative blowing up. These facts highlight the emergence of a finite big freeze singularity in the past. \footnote{Please see Section \ref{introduction} for the definition of a big freeze singularity and the classification of the other cosmological singularities related to dark energy. In our case this singular event happens at a finite past of the Universe.}

As a summary, we find that the original loitering effect for positive $\kappa$ can be substituted by a point with a minimum scale factor $a_m$, where a bounce ($H^2=0$ and $dH^2/d\rho$ remains finite), a past quasi-sudden singularity ($H^2=0$ and $dH^2/d\rho$ nearly diverges) or a past big freeze singularity ($H^2$ and $dH^2/d\rho$ diverge) may emerge.

\subsubsection{$\beta<0$}

\begin{figure}[!h]
\graphicspath{{fig/}}
\includegraphics[scale=1.0]{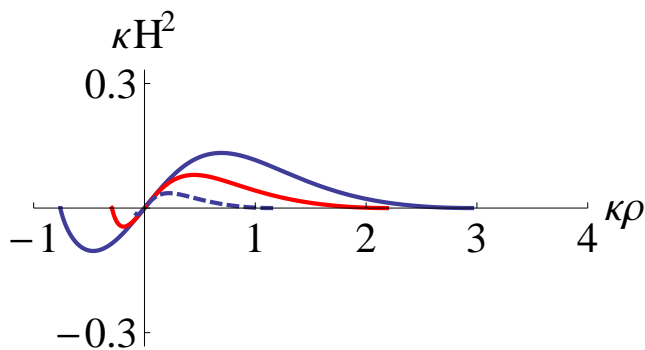}
\caption{Graphical representation of $\kappa H^2$ as a function of $\kappa\rho$ for a radiation dominated universe ($w=1/3$). Different curves represent different $\beta$. The solid blue, red, and dashed blue curves show the representations for $\beta=-10^{-2}$, $-1/10$, and $-3/10$, respectively.}
\label{rad3}
\end{figure}

In FIG.~\ref{rad3}, we show the representations of $\kappa H^2$ as a function of $\kappa \rho$ for $\beta<0$. We find that, unlike what we concluded previously, the loitering effects ($H^2\propto{\delta\rho}^2$) and the bouncing solutions ($H^2\propto{\delta\rho}$) are robust against the decrease of $\beta$ below zero. Furthermore, we also find that the smaller the value of $\beta$, the smaller the value of $|\kappa\rho|$ at the loitering event or the bounce.

\subsubsection{$\beta>1/4$}

\begin{figure}[!h]
\graphicspath{{fig/}}
\includegraphics[scale=0.8]{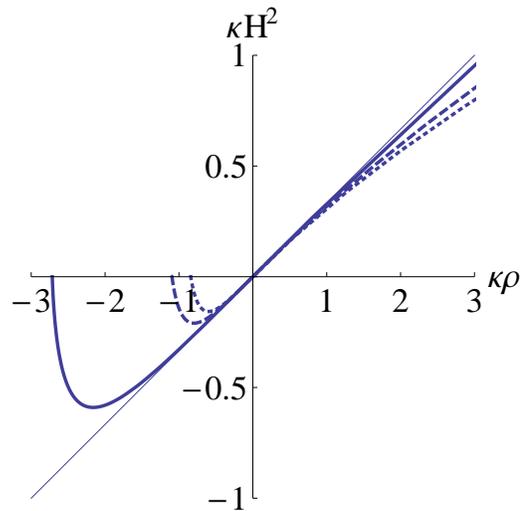}
\caption{Graphical representation of $\kappa H^2$ as a function of $\kappa\rho$ for a radiation dominated universe ($w=1/3$) with $\beta>1/4$. The solid, dashed, and dotted curves represent the solutions for $\beta=3/10$, $7/20$, and $37/100$, respectively. Note that the solution $H^2=\rho/3$, which is the solution within $R+R^2$ gravity ($\beta=1/4$), is also shown in this figure.}
\label{rad4}
\end{figure}

\begin{figure}[!h]
\graphicspath{{fig/}}
\includegraphics[scale=0.8]{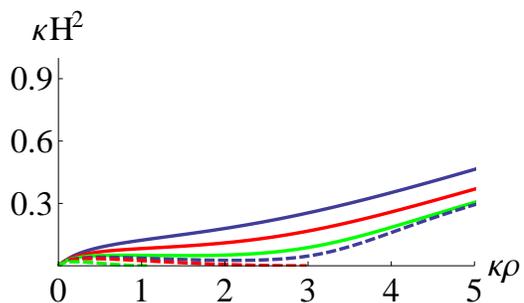}
\caption{Graphical representation of $\kappa H^2$ as a function of $\kappa\rho$ for a radiation dominated universe ($w=1/3$) with $\beta>1/4$ and $\kappa$ being positive. The solid blue, red, and green curves represent the solutions for $\beta=7/10$, $4/5$, and $9/10$, respectively. The dashed blue, red and green curves represent the solutions for $\beta=19/20$, $\beta=1$ and $\beta=6/5$, respectively.}
\label{rad5}
\end{figure}

On the other hand, in FIG.~\ref{rad4} and FIG.~\ref{rad5} we show the representations of $\kappa H^2$ as a function of $\kappa\rho$ for $\beta>1/4$ (or $\alpha<0$) in a radiation dominated universe. The straight line $H^2=\rho/3$, which represents the solution within $R+R^2$ gravity ($\beta=1/4$), is also exhibited. One can see that the bouncing solutions for negative $\kappa$ are robust against a change of $\beta$. However, the loitering solutions for positive $\kappa$ in the EiBI theory become big bang solutions where $H^2\propto\rho$ for large $\rho$. These solutions converge to $H^2=\rho/3$ when the value of $\beta$ approaches $\beta\approx 1/4$.

Interestingly, we also find from the dashed blue and solid green curves in FIG.~\ref{rad5} that for larger values of $\beta$, there could be a plateau in the $H^2$ function for positive $\kappa$, for a radiation dominated universe. This stage may correspond to a de Sitter inflationary expansion phase after the big bang singularity. This inflationary phase is then followed by a classical expansion described well in the context of GR. Furthermore, when $\beta\ge 1$, the solutions with a loitering effect are again recovered (see the dashed red and dashed green curves in FIG.~\ref{rad5}).

Before concluding, we notice that because this theory reduces to GR at the low energy limit, all the radiation dominated universe will be asymptotically flat at that limit.
 
\section{conclusions}

\begin{table*}
 \begin{center}
  \begin{tabular}{||c||c||c||}
  \hline 
    & $\kappa>0$ & $\kappa<0$ \\
  \hline\hline
   $\beta<0$ &  &  \\ 
  \cline{1-1}
   $\beta=0$ & past loitering effect& bounce \\
   (EiBI theory)& & \\
  \hline\ 
   $0<\beta<\beta_{\star}$ & bounce &  \\
   \cline{1-2}
   $\beta=\beta_{\star}$&past quasi-sudden singularity & \\
   \cline{1-2}
   $\beta_{\star}<\beta<1/4$& past big freeze singularity& big bang singularity\\
  \cline{1-2}
   $\beta=1/4$ && \\
   Palatini $R^2$ theory&big bang singularity&\\
   \cline{1-1}\cline{3-3}
   $1/4<\beta<1$&&\\
   \cline{1-2}
   $\beta\lesssim 1$&big bang singularity$+$de Sitter&bounce\\
   \cline{1-2}
   $\beta\geq 1$& past loitering effect&\\
  \hline
  \end{tabular}
  \caption{This table summarizes how the big bang singularity in GR is altered in the modified EiBI theory for a radiation dominated universe. If $\kappa<0$, the big bang is substituted by a bounce except for the regions of the parameter space $0<\beta\leq1/4$ where the big bang is still present. If $\kappa>0$, the big bang singularity can be altered by a loitering effect, a bounce, what we named a quasi-sudden singularity, or a big freeze singularity in the past. However, for $1/4\leq\beta<1$, the big bang singularity exists. Furthermore, the big bang singularity may be followed by a de Sitter inflationary stage for $\beta\lesssim 1$.}
    \label{summary}
 \end{center}
\end{table*} 

Since the proposal of the Born-Infeld action for classical electrodynamics, modified theories of gravity inspired on such a proposal and with an elegant determinantal structure in their actions have been widely investigated (see Refs.~\cite{Comelli:2004qr,Vollick:2003qp,Vollick:2005gc,Wohlfarth:2003ss,Nieto:2004qj,Comelli:2005tn}). Despite the large amount of works in this subject, the very interesting generalization through the addition of a pure trace term into the gravitational Lagrangian in the Palatini formalism has not been considered before. This modification gives rise to the most general action constructed from a rank two tensor that contains up to first order terms in the curvature. Such a theory is expected to not only preserve the great achievement of GR at low energies, but also to generate more drastic deviations from GR than those accomplished within the original Born-Infeld-inspired theories at high energies. Modified theories with this term have only been investigated in the pure metric formalism \cite{Comelli:2004qr} and in the teleparallel representation \cite{Fiorini:2013kba}. The former inevitably suffers from troublesome fourth order field equations for the metric or from ghost instabilities \cite{Deser:1998rj}, which suggests the need of some alternative approach to overcome these problems. The latter, which flees from the ghosts and results in second order field equations, leads in most of the cases to the substitution of the big bang by smoother cosmological singularities \cite{Bouhmadi-Lopez:2014tna} or a de Sitter inflationary stage \cite{Fiorini:2013kba}.

Inspired by these motivations, in this paper we generalize the EiBI theory, which is formulated within the Palatini formalism, by adding a pure trace term into the determinantal Lagrangian, and analyze the cosmological solutions of this theory by assuming a homogeneous and isotropic universe for its largest scale. As we expect, the early cosmological expansion to be modified as compared with GR or EiBI theory, we assume that the Universe is filled with radiation. Following a similar approach to that proposed in Ref.~\cite{Makarenko:2014lxa}, the behaviors of the cosmological solutions are analyzed using a parametric Friedmann equation.

As a summary, we find that if $\kappa<0$, the big bang is substituted by a bounce except for the regions of the parameter space $0<\beta\leq1/4$ where the big bang singularity exists. Note that in Refs.~\cite{Makarenko:2014lxa,Odintsov:2014yaa,Makarenko:2014cca} the authors showed that the bouncing solutions in the EiBI theory are robust against the changes of the Lagrangian through an additional $f(R)$ term or some functional extensions. On the other hand, if $\kappa>0$, we find that the big bang singularity can be altered by a loitering effect ($\beta\leq 0$ or $\beta\geq 1$), a bounce ($0<\beta<\beta_{\star}$), what we named a quasi-sudden singularity ($\beta=\beta_{\star}$), or a big freeze singularity in the past ($\beta_{\star}<\beta<1/4$). However, for $1/4\leq\beta<1$, the big bang singularity remains. Most interestingly, the big bang singularity may be followed by a de Sitter inflationary stage for $\beta\lesssim 1$. This can be verified by the plateau in the $H^2$ function as shown in FIG.~\ref{rad5}. The inflationary phase is superseded by a standard cosmological expansion. We summarizes our results in TABLE~\ref{summary}. Moreover, we should emphasize that the cosmological solutions that emerge in this theory are all stemmed from pure geometrical effects. Only a radiation dominated universe is assumed and there is no need of any additional fields or exotic matters to drive these cosmological solutions.

\acknowledgments

M.B.L. is supported by the Portuguese Agency ``Funda\c{c}\~{a}o para a Ci\^{e}ncia e Tecnologia" through an
Investigador FCT Research contract, with reference IF/01442/2013/CP1196/CT0001. She also wishes to acknowledge the hospitality of LeCosPA Center at the National Taiwan University during the completion of part of this work and the support from the Portuguese Grants PTDC/FIS/111032/2009 and UID/MAT/00212/2013  and the partial support from the Basque government Grant No. IT592-13 (Spain).
C.-Y.C. and P.C. are supported by Taiwan National Science Council under Project No. NSC 97-2112-M-002-026-MY3 and by Taiwan’s National Center for Theoretical Sciences (NCTS). P.C. is in addition supported by US Department of Energy under Contract No. DE-AC03-76SF00515.


\begin{thebibliography}{99}

\bibitem{gravitation} 
  C.~W.~Misner, K.~S.~Thorne, and J.~A.~Wheeler, \textit{Gravitation}, (W.~H.~Freeman, 1973).

\bibitem{largescale} 
  S.~W.~Hawking, G.~F.~R.~Ellis, \textit{The Large Scale Structure of Space-Time}, (Cambridge University Press, 1973).

\bibitem{Born:1934gh}
  M.~Born and L.~Infeld,
  Proc.\ Roy.\ Soc.\ Lond.\ A {\bf 144} (1934) 425.

\bibitem{Deser:1998rj}
  S.~Deser and G.~W.~Gibbons,
  Class.\ Quant.\ Grav.\  {\bf 15} (1998) L35.

\bibitem{Comelli:2004qr}
  D.~Comelli and A.~Dolgov,
  JHEP {\bf 0411} (2004) 062.

\bibitem{Wohlfarth:2003ss}
  M.~N.~R.~Wohlfarth,
  Class.\ Quant.\ Grav.\  {\bf 21} (2004) 1927
   [Class.\ Quant.\ Grav.\  {\bf 21} (2004) 5297].

\bibitem{Nieto:2004qj}
  J.~A.~Nieto,
  Phys.\ Rev.\ D {\bf 70} (2004) 044042.

\bibitem{Comelli:2005tn}
  D.~Comelli,
  Phys.\ Rev.\ D {\bf 72} (2005) 064018.

\bibitem{Vollick:2003qp}
  D.~N.~Vollick,
  Phys.\ Rev.\ D {\bf 69} (2004) 064030.

\bibitem{Vollick:2005gc}
  D.~N.~Vollick,
  Phys.\ Rev.\ D {\bf 72} (2005) 084026.

\bibitem{Ferraro:2006jd}
  R.~Ferraro and F.~Fiorini,
  Phys.\ Rev.\ D {\bf 75} (2007) 084031.

\bibitem{Ferraro:2008ey}
  R.~Ferraro and F.~Fiorini,
  Phys.\ Rev.\ D {\bf 78} (2008) 124019.
  
\bibitem{Fiorini:2009ux}
  F.~Fiorini and R.~Ferraro,
  Int.\ J.\ Mod.\ Phys.\ A {\bf 24} (2009) 1686.
  
\bibitem{Ferraro:2009zk}
  R.~Ferraro and F.~Fiorini,
  Phys.\ Lett.\ B {\bf 692} (2010) 206.

\bibitem{Fiorini:2013kba}
  F.~Fiorini,
  Phys.\ Rev.\ Lett.\  {\bf 111} (2013) 4,  041104.


\bibitem{Banados:2010ix}
  M.~Ba\~{n}ados and P.~G.~Ferreira,
  Phys.\ Rev.\ Lett.\  {\bf 105} (2010) 011101
   [Phys.\ Rev.\ Lett.\  {\bf 113} (2014) 11,  119901].


\bibitem{Scargill:2012kg}
  J.~H.~C.~Scargill, M.~Ba\~{n}ados and P.~G.~Ferreira,
  Phys.\ Rev.\ D {\bf 86} (2012) 103533.

\bibitem{Pani:2011mg}
  P.~Pani, V.~Cardoso and T.~Delsate,
  Phys.\ Rev.\ Lett.\  {\bf 107} (2011) 031101.

\bibitem{Bouhmadi-Lopez:2014jfa}
  M.~Bouhmadi-L\'{o}pez, C.~Y.~Chen and P.~Chen,
  Eur.\ Phys.\ J.\ C {\bf 75} (2015) 2,  90.
  
 
\bibitem{Bouhmadi-Lopez:2013lha}
  M.~Bouhmadi-L\'{o}pez, C.~Y.~Chen and P.~Chen,
  Eur.\ Phys.\ J.\ C {\bf 74} (2014) 2802.

\bibitem{Potapov:2014iva}
  A.~A.~Potapov, R.~Izmailov, O.~Mikolaychuk, N.~Mikolaychuk, M.~Ghosh and K.~K.~Nandi,
  JCAP {\bf 1507} (2015) 07,  018.

\bibitem{Izmailov:2015xsa}
  R.~Izmailov, A.~A.~Potapov, A.~I.~Filippov, M.~Ghosh and K.~K.~Nandi,
  Mod.\ Phys.\ Lett.\ A {\bf 30} (2015) 11,  1550056.

\bibitem{Tamang:2015tmd}
  A.~Tamang, A.~A.~Potapov, R.~Lukmanova, R.~Izmailov and K.~K.~Nandi,
  Class.\ Quant.\ Grav.\  {\bf 32} (2015) 23,  235028.

\bibitem{Avelino:2012ue}
  P.~P.~Avelino and R.~Z.~Ferreira,
  Phys.\ Rev.\ D {\bf 86} (2012) 041501.


\bibitem{EscamillaRivera:2012vz}
  C.~Escamilla-Rivera, M.~Ba\~{n}ados and P.~G.~Ferreira,
  Phys.\ Rev.\ D {\bf 85} (2012) 087302.

\bibitem{Yang:2013hsa}
  K.~Yang, X.~L.~Du and Y.~X.~Liu,
  Phys.\ Rev.\ D {\bf 88} (2013) 12,  124037.

\bibitem{Du:2014jka}
  X.~L.~Du, K.~Yang, X.~H.~Meng and Y.~X.~Liu,
  Phys.\ Rev.\ D {\bf 90} (2014) 4,  044054.

\bibitem{Wei:2014dka}
  S.~W.~Wei, K.~Yang and Y.~X.~Liu,
  Eur.\ Phys.\ J.\ C {\bf 75} (2015) 6,  253.

\bibitem{Delsate:2012ky}
  T.~Delsate and J.~Steinhoff,
  Phys.\ Rev.\ Lett.\  {\bf 109} (2012) 021101.

\bibitem{Pani:2012qb}
  P.~Pani, T.~Delsate and V.~Cardoso,
  Phys.\ Rev.\ D {\bf 85} (2012) 084020.

\bibitem{Casanellas:2011kf}
  J.~Casanellas, P.~Pani, I.~Lopes and V.~Cardoso,
  Astrophys.\ J.\  {\bf 745} (2012) 15.

\bibitem{Avelino:2012ge}
  P.~P.~Avelino,
  Phys.\ Rev.\ D {\bf 85} (2012) 104053.

\bibitem{Avelino:2012qe}
  P.~P.~Avelino,
  JCAP {\bf 1211} (2012) 022.

\bibitem{Harko:2013wka}
  T.~Harko, F.~S.~N.~Lobo, M.~K.~Mak and S.~V.~Sushkov,
  Phys.\ Rev.\ D {\bf 88} (2013) 044032.

\bibitem{Harko:2013aya}
  T.~Harko, F.~S.~N.~Lobo, M.~K.~Mak and S.~V.~Sushkov,
  Mod.\ Phys.\ Lett.\ A {\bf 30} (2015) 35,  1550190.

\bibitem{Sham:2013cya}
  Y.~H.~Sham, L.~M.~Lin and P.~T.~Leung,
  Astrophys.\ J.\  {\bf 781} (2014) 2,  66.
  
\bibitem{Makarenko:2014lxa}
  A.~N.~Makarenko, S.~Odintsov and G.~J.~Olmo,
  Phys.\ Rev.\ D {\bf 90} (2014) 024066.

\bibitem{Odintsov:2014yaa}
  S.~D.~Odintsov, G.~J.~Olmo and D.~Rubiera-Garcia,
  Phys.\ Rev.\ D {\bf 90} (2014) 4,  044003.

\bibitem{Makarenko:2014cca}
  A.~N.~Makarenko, S.~D.~Odintsov, G.~J.~Olmo and D.~Rubiera-Garcia,
  TSPU Bulletin {\bf 12} (2014) 158.

\bibitem{Makarenko:2014nca}
  A.~N.~Makarenko, S.~D.~Odintsov and G.~J.~Olmo,
  Phys.\ Lett.\ B {\bf 734} (2014) 36.

\bibitem{Pani:2012qd}
  P.~Pani and T.~P.~Sotiriou,
  Phys.\ Rev.\ Lett.\  {\bf 109} (2012) 251102.

\bibitem{Shaikh:2015oha}
  R.~Shaikh,
  Phys.\ Rev.\ D {\bf 92} (2015) 024015.

\bibitem{Jana:2015cha}
  S.~Jana and S.~Kar,
  Phys.\ Rev.\ D {\bf 92} (2015) 084004.
  
\bibitem{Sotani:2015tya}
  H.~Sotani,
  Phys.\ Rev.\ D {\bf 91} (2015) 8,  084020.
  
\bibitem{Cho:2014xaa}
  I.~Cho and N.~K.~Singh,
  Eur.\ Phys.\ J.\ C {\bf 75} (2015) 6,  240.
  
\bibitem{Sotani:2014lua}
  H.~Sotani and U.~Miyamoto,
  Phys.\ Rev.\ D {\bf 90} (2014) 12,  124087.

\bibitem{Bouhmadi-Lopez:2014tna}
  M.~Bouhmadi-L\'{o}pez, C.~Y.~Chen and P.~Chen,
  Phys.\ Rev.\ D {\bf 90} (2014) 12,  123518.

\bibitem{Cho:2012vg}
  I.~Cho, H.~C.~Kim and T.~Moon,
  Phys.\ Rev.\ D {\bf 86} (2012) 084018.

\bibitem{Cho:2013pea}
  I.~Cho, H.~C.~Kim and T.~Moon,
  Phys.\ Rev.\ Lett.\  {\bf 111} (2013) 071301.

\bibitem{Cho:2013usa}
  I.~Cho and H.~C.~Kim,
  Phys.\ Rev.\ D {\bf 88} (2013) 064038.

\bibitem{Jimenez:2014fla}
  J.~B.~Jim\'{e}nez, L.~Heisenberg and G.~J.~Olmo,
  JCAP {\bf 1411} (2014) 004.

\bibitem{Jimenez:2015caa}
  J.~B.~Jim\'{e}nez, L.~Heisenberg and G.~J.~Olmo,
  JCAP {\bf 1506} (2015) 026.

  \bibitem{tele} 
  R.~Aldrovandi, and J.~G.~Pereira, \textit{Teleparallel Gravity: An Introduction}, (Springer, Dordrecht, 2012).

\bibitem{Nojiri:2005sx}
  S.~Nojiri, S.~D.~Odintsov and S.~Tsujikawa,
  Phys.\ Rev.\ D {\bf 71} (2005) 063004.
  
\bibitem{FernandezJambrina:2004yy}
  L.~Fern\'{a}ndez-Jambrina and R.~Lazkoz,
  Phys.\ Rev.\ D {\bf 70} (2004) 121503.
  
\bibitem{FernandezJambrina:2006hj}
  L.~Fern\'{a}ndez-Jambrina and R.~Lazkoz,
  Phys.\ Rev.\ D {\bf 74} (2006) 064030.

\bibitem{Fernandez-Jambrina:2014sga}
  L.~Fern\'{a}ndez-Jambrina,
  Phys.\ Rev.\ D {\bf 90} (2014) 6,  064014.

\bibitem{Starobinsky:1999yw}
  A.~A.~Starobinsky,
  Grav.\ Cosmol.\  {\bf 6} (2000) 157.

\bibitem{Caldwell:2003vq}
  R.~R.~Caldwell, M.~Kamionkowski and N.~N.~Weinberg,
  Phys.\ Rev.\ Lett.\  {\bf 91} (2003) 071301.

\bibitem{Caldwell:1999ew}
  R.~R.~Caldwell,
  Phys.\ Lett.\ B {\bf 545} (2002) 23.

\bibitem{Carroll:2003st}
  S.~M.~Carroll, M.~Hoffman and M.~Trodden,
  Phys.\ Rev.\ D {\bf 68} (2003) 023509.
  
\bibitem{Chimento:2003qy}
  L.~P.~Chimento and R.~Lazkoz,
  Phys.\ Rev.\ Lett.\  {\bf 91} (2003) 211301.
  
\bibitem{Dabrowski:2003jm}
  M.~P.~D\c{a}browski, T.~Stachowiak and M.~Szyd\a{l}owski,
  Phys.\ Rev.\ D {\bf 68} (2003) 103519.


\bibitem{GonzalezDiaz:2003rf}
  P.~F.~Gonz\'{a}lez-D\'{i}az,
  Phys.\ Lett.\ B {\bf 586} (2004) 1.

\bibitem{GonzalezDiaz:2004vq}
  P.~F.~Gonz\'{a}lez-D\'{i}az,
  Phys.\ Rev.\ D {\bf 69} (2004) 063522.

\bibitem{Gorini:2003wa}
  V.~Gorini, A.~Y.~Kamenshchik, U.~Moschella and V.~Pasquier,
  Phys.\ Rev.\ D {\bf 69} (2004) 123512.

\bibitem{Barrow:2004xh}
  J.~D.~Barrow,
  Class.\ Quant.\ Grav.\  {\bf 21} (2004) L79.

\bibitem{Nojiri:2004pf}
  S.~Nojiri and S.~D.~Odintsov,
  Phys.\ Rev.\ D {\bf 70} (2004) 103522.

\bibitem{BouhmadiLopez:2007qb}
  M.~Bouhmadi-L\'{o}pez, P.~F.~Gonz\'{a}lez-D\'{i}az and P.~Mart\'{i}n-Moruno,
  Int.\ J.\ Mod.\ Phys.\ D {\bf 17} (2008) 2269.

\bibitem{BouhmadiLopez:2006fu}
  M.~Bouhmadi-L\'{o}pez, P.~F.~Gonz\'{a}lez-D\'{i}az and P.~Mart\'{i}n-Moruno,
  Phys.\ Lett.\ B {\bf 659} (2008) 1.

\bibitem{Nojiri:2005sr}
  S.~Nojiri and S.~D.~Odintsov,
  Phys.\ Rev.\ D {\bf 72} (2005) 023003.

\bibitem{Nojiri:2008fk}
  S.~Nojiri and S.~D.~Odintsov,
  Phys.\ Rev.\ D {\bf 78} (2008) 046006.
  
\bibitem{Bamba:2008ut}
  K.~Bamba, S.~Nojiri and S.~D.~Odintsov,
  JCAP {\bf 0810} (2008) 045.

\bibitem{Ruzmaikina1970}
  T. Ruzmaikina and A. A. Ruzmaikin, Sov. Phys. JETP {\bf 30} (1970) 372.

\bibitem{Stefancic:2004kb}
  H.~\v{S}tefan\v{c}i\'{c},
  Phys.\ Rev.\ D {\bf 71} (2005) 084024.

\bibitem{BouhmadiLopez:2005gk}
  M.~Bouhmadi-L\'{o}pez,
  Nucl.\ Phys.\ B {\bf 797} (2008) 78.

\bibitem{Frampton:2011sp}
  P.~H.~Frampton, K.~J.~Ludwick and R.~J.~Scherrer,
  Phys.\ Rev.\ D {\bf 84} (2011) 063003.

\bibitem{Brevik:2011mm}
  I.~Brevik, E.~Elizalde, S.~Nojiri and S.~D.~Odintsov,
  Phys.\ Rev.\ D {\bf 84} (2011) 103508.

\bibitem{Bouhmadi-Lopez:2013nma}
  M.~Bouhmadi-L\'{o}pez, P.~Chen and Y.~-W.~Liu,
  Eur.\ Phys.\ J.\ C {\bf 73} (2013) 9,  2546.

\bibitem{Bouhmadi-Lopez:2014cca} 
  M.~Bouhmadi-L\'{o}pez, A.~Errahmani, P.~Mart\'{\i}n-Moruno, T.~Ouali and Y.~Tavakoli,
Int.\ J.\ Mod.\ Phys.\ D {\bf 24} (2015) 1550078; I.~Albarran, M.~Bouhmadi-L\'{o}pez, F.~Cabral and P.~Mart\'{\i}n-Moruno,
  JCAP {\bf 1511} (2015) 11,  044.
  
\bibitem{Meng:2003bk}
  X.~H.~Meng and P.~Wang,
  Gen.\ Rel.\ Grav.\  {\bf 36} (2004) 2673.

\end{thebibliography}
\end{document}